\documentclass[aps,prl,twocolumn,citeautoscript,superscriptaddress]{revtex4-1} 
\usepackage{amsmath,amssymb,mathrsfs,bm,feynmf,setspace}
\usepackage{natbib}
\usepackage{graphicx}
\usepackage{color} 
\usepackage{hyperref} 

\newcommand{\al}{\alpha}

\newcommand{\s}{\sigma}

\newcommand{\w}{\omega}

\newcommand{\De}{\Delta} 
\newcommand{\G}{\Gamma}

\newcommand{\D}{\Delta}

\newcommand{\ket}[1]{\left| #1\right\rangle}

\newcommand{\beq}{\begin{equation}}
\newcommand{\eeq}{\end{equation}}
\newcommand{\Beq}{\begin{eqnarray}}
\newcommand{\Eeq}{\end{eqnarray}}
\newcommand{\bml}{\begin{multline}}

\newcommand{\eeqm}{\end{multline}}

\newcommand{\bsp}{\begin{split}}
\newcommand{\esp}{\end{split}}

\renewcommand{\b}[1]{{\mathbf #1}}

\newcommand{\inv}{^{-1}}

\newcommand{\req}[1]{Eq.~(\ref{eq:#1})}
\newcommand{\Req}[1]{Equation~(\ref{eq:#1})}

\newcommand{\R}{\mathbf{R}}
\newcommand{\wn}{\omega_n}
\newcommand{\methods}{Supplementary Material}

\begin{document}
\title{Correlation effects on 3D topological phases: from bulk to boundary}
\author{Ara \surname{Go}}
\affiliation{Department of Physics and Astronomy and Center for Theoretical Physics, Seoul National University, Seoul 151-747, Korea}
\affiliation{Department of Physics, Ewha Womans University, Seoul 120-750, Korea}
\author{William \surname{Witczak-Krempa}}
\affiliation{Department of Physics, University of Toronto, Toronto, Ontario M5S 1A7, Canada}
\author{Gun Sang \surname{Jeon}}
\affiliation{Department of Physics, Ewha Womans University, Seoul 120-750, Korea}
\author{Kwon \surname{Park}}
\affiliation{School of Physics, Korea Institute for Advanced Study, Seoul 130-722, Korea}
\author{Yong Baek \surname{Kim}}
\affiliation{Department of Physics, University of Toronto, Toronto, Ontario M5S 1A7, Canada}
\affiliation{School of Physics, Korea Institute for Advanced Study, Seoul 130-722, Korea}
\date{\today}

\begin{abstract}
Topological phases of quantum matter defy characterization by conventional order parameters
but can exhibit quantized electro-magnetic response and/or protected surface
states. We examine such phenomena in a model for three-dimensional correlated 
complex oxides, the pyrochlore iridates. The model realizes interacting
topological insulators with and without time-reversal
symmetry, and topological Weyl semimetals. We use cellular dynamical mean
field theory, a method that incorporates
quantum-many-body effects and allows us to evaluate the magneto-electric
topological response coefficient in correlated systems. This invariant is used
to unravel the presence of an interacting axion insulator absent within a simple mean field study. 
We corroborate 
our bulk results by studying the evolution 
of the topological boundary states in the presence of interactions.
Consequences for experiments and for
the search for correlated materials with symmetry-protected topological order are given.
\end{abstract}
\maketitle
The interplay of symmetry and topology has recently proven a rich avenue
for the discovery of new phases of matter, some of which have been experimentally
identified. A recent example is the prediction and subsequent observation of 
topological insulators (TIs) preserving time-reversal symmetry (TRS)~\cite{Hasan2010,Qi2011}.
The symmetry-protected
topological orders in such phases cannot be fully characterized by conventional order parameters. 
They can, however, display universal electromagnetic
response and/or robust boundary states. 
This follows from the presence of non-trivial quantum entanglement in the ground state of these  
phases~\cite{Turner2010:2}.
In TRS-protected topological band insulators, such order can be characterized by
a topological invariant in terms of the single-particle wavefunction\cite{Hasan2010}. This approach cannot
be applied in the presence of interactions. Rather, one can 
ask if the correlated material exhibits a universal and quantized physical response to a given external 
perturbation. 
For three-dimensional (3D) TIs, the quantized response is
the magneto-electric effect \cite{Hughes2008}, according to which an externally applied
electric field on the sample generates a parallel magnetic field, and vice versa, with 
the response coefficient depending solely on universal constants. It was argued on the grounds of topological
field theory that the magneto-electric effect remains a well-defined topological response
in the presence of interactions \cite{Wang2010}. Moreover,
a topological index was given in terms of the full interacting electronic Green's function.
As some of the considerations in the establishment of such an invariant are rather abstract,
it would be desirable to have a concrete verification of such an important claim.

We calculate this topological invariant within a strongly correlated electronic Hamiltonian
by means of cellular dynamical mean-field theory (CDMFT)~\cite{Kotliar2001,Maier2005}, which
gives
access to the Green's function of the interacting electrons. This index allows us to 
determine the presence of correlated topological insulators, with and without TRS
(here, the latter case corresponds to an axion insulator), and their breakdown for sufficiently
large correlations.

A complementary aspect of the quantized magneto-electric effect is the presence of protected
surface states, which we verify
by analysing our interacting model in a finite-slab
geometry. We do find that the TI surface states are robust to interactions,
thus establishing the bulk-boundary connection for an interacting system and providing
a check for the non-trivial topological index.
Further, the surface state analysis allows us to study correlation effects on a closely related gapless phase:
the topological Weyl semimetal, wich has Weyl-fermion
excitations and non-trivial surface states~\cite{Wan2011,Balents2011,Burkov2011,Yang2011}.

We use a model relevant to a class of 3D complex oxides, the pyrochlore 
iridates\cite{Pesin2010,Wan2011,Yang2010,Kargarian2011,Witczak-Krempa2011}. 
These materials, and closely related Iridium-based compounds, are currently under close
experimental scrutiny due to recent proposals for topological 
phases~\cite{Pesin2010,tmi,Wan2011,Yang2010,Kargarian2011,Witczak-Krempa2011}.
As correlations
seem important in these $d$-electron compounds, our work can be instrumental
in their analysis. However, we emphasize that as we are dealing with topological phases, many of our
results are expected to hold in general. Indeed, we envision that our methods can be fruitfully 
used in the analysis of correlated symmetry-protected topological
ordered states and combined with \emph{ab initio} tools in the quest for experimentally 
relevant candidate materials~\cite{Kotliar2006}.

\emph{Model:} The pyrochlore iridates, $R_2$Ir$_2$O$_7$, are 3D complex oxides where $R$ is Yttrium or a rare earth.
In many instances $R$ is non-magnetic and the physics is mainly dictated by Iridium's (Ir) $5d$-electrons.
Due to the larger extent of the $5d$ atomic orbitals (compared with $3d$), the energy scales associated
with spin-orbit coupling and local repulsion are comparable. This sets the stage for 
the interplay between
band topology and Mott physics.
A microscopically-tailored model that captures this interplay is the following Hubbard Hamiltonian for the $5d$-electrons 
hopping on the Iridium pyrochlore lattice 
with onsite Coulomb repulsion~\cite{Witczak-Krempa2011}:
\begin{align}\label{eq:H}
  H = \sum_{\langle \R i, \R^\prime i^\prime \rangle,\sigma\sigma^\prime}
  ([T_o]^{ii^\prime}_{\sigma\sigma^\prime}
  +[T_d]^{ii^\prime}_{\sigma\sigma^\prime})
  c^\dagger_{\R i \sigma}
  c^{}_{\R^\prime i^\prime \sigma^\prime} \nonumber\\
  -\mu \sum_{\R i, \sigma} 
  c^\dagger_{\R i \sigma}
  c^{}_{\R i \sigma}
  +U \sum_{\R i} n^{}_{\R i \uparrow} n^{}_{\R i \downarrow} \;,
\end{align}
where $c^{}_{\R i \sigma}$ annihilates an electron
with pseudospin $\sigma$ at the $i$th basis site of the Bravais lattice vector $\R$.
The index $i$ runs from 1 to 4 and labels the corners of a tetrahedron.
The hopping matrix $T_o$ arises from oxygen-mediated hopping between the Ir atoms \cite{Pesin2010}
with amplitude $t$, while $T_d$ from the Ir-Ir hopping due to the
\emph{direct} overlap between the extended $5d$-orbitals. The latter depends on two
energy scales, $t_\sigma$ and $t_\pi$,
arising from the $\sigma$- and $\pi$-bonding between the orbitals, respectively.
The chemical potential, $\mu$, is such that each Ir atom contributes 
a single 
pseudospin-$1/2$ electron. 
The pseudospin arises from the combined effect of crystal fields and spin-orbit coupling~\cite{Kim2009}.
Finally, the Hubbard repulsion $U$ generates correlations
by penalizing double occupation and thus drives the system away from simple single-particle
physics. (We shall use the oxygen-mediated hopping amplitude, $t$, as our comparison scale.)

The phase diagram of the above Hamiltonian was previously analyzed by treating the onsite
repulsion within a mean-field Hartree-Fock (HF) approach~\cite{Witczak-Krempa2011}, 
which allows for a single-particle
description. It was found that for small $U/t$, one obtains topological
insulator and metallic phases, depending on the ratios $t_\s/t$ and $t_\pi/t$.
At sufficiently large $U$, the systems become magnetic. Near the magnetic transitions,
it was found that topological Weyl semimetals (TWS) arise.
Here, we shall focus on a representative set of hopping parameters: $t_\sigma/t=1$,
with the ratio $t_\pi/t_\sigma = -2/3$ fixed. 
In that case, the HF mean-field theory predicts that the system undergoes successive transitions from a TI to a  
TWS, and to an antiferromagnetic insulator (AFI) as one increases $U$. It is worth noting that the same succession of phases 
can be found within the HF framework for $t_\s/t<-1.67$, and we thus expect that the results we present below can
be applied there as well. A detailed study of the full phase diagram is left for future work.

We use the above model to examine the fate of these phases and transitions within CDMFT. 
This method has been widely used to investigate correlated microscopic models~\cite{Maier2005}
but only recently was it applied to topological phases~\cite{Wu2011}, specializing to two dimensions.  
We emphasize that CDMFT fully incorporates the quantum many-body effects within
a cluster (unit cell here).

\begin{figure}[t]
	\includegraphics[width=0.85\linewidth]{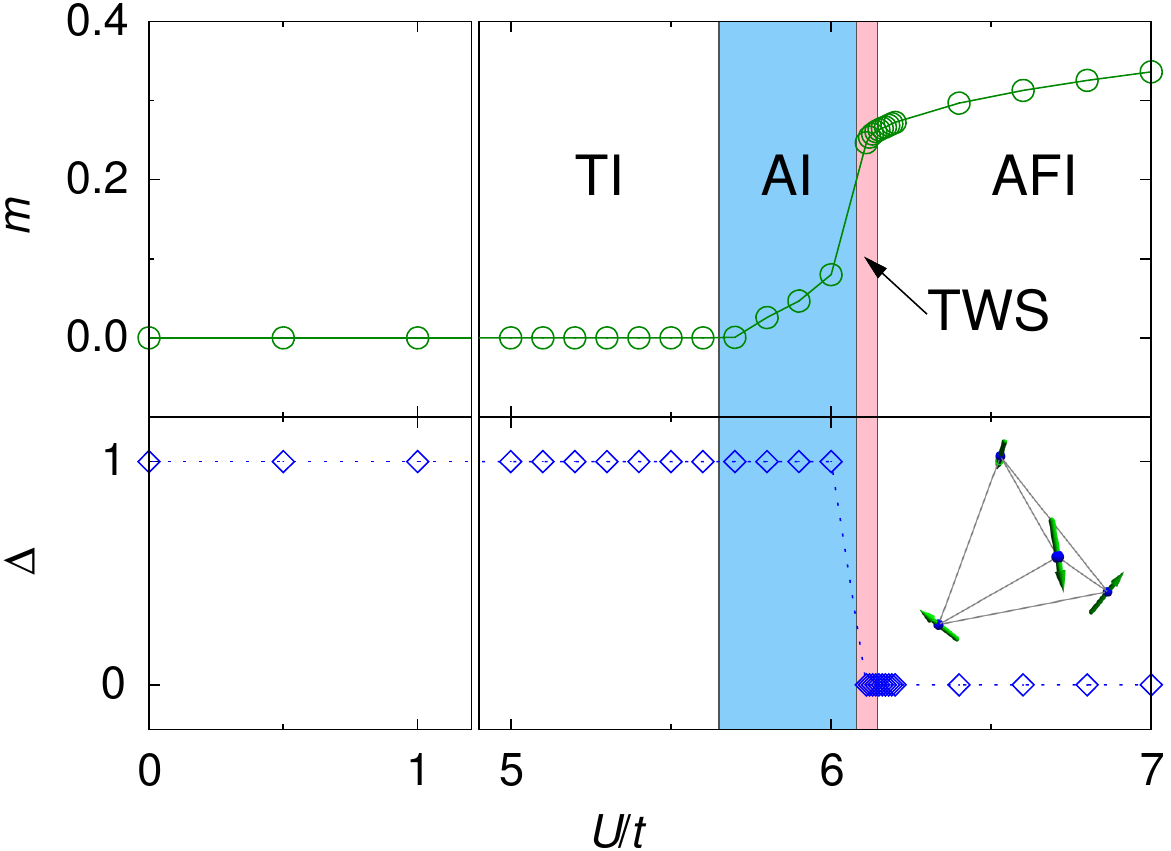}
	\caption{\label{fig.phase}%
	Magnetization ($m$) and topological index ($\De$) versus interaction strength.
	An interaction-driven topological transition accompanies an abrupt change of the magnetization.
	In the intermediate region,
	a topologically nontrivial insulator with a finite magnetization
	indicates the realization of an interacting axion insulator (AI).
	As the interaction strength increases,
        a topological Weyl semimetal (TWS) appears after the magnetization jump.
	At large $U$, the system is a topologically-trivial antiferromagnetic insulator (AFI). 
	The magnetic structure is illustrated in the inset.
	}
\end{figure}

\begin{figure}[tb]
	\includegraphics[width=1.02\linewidth]{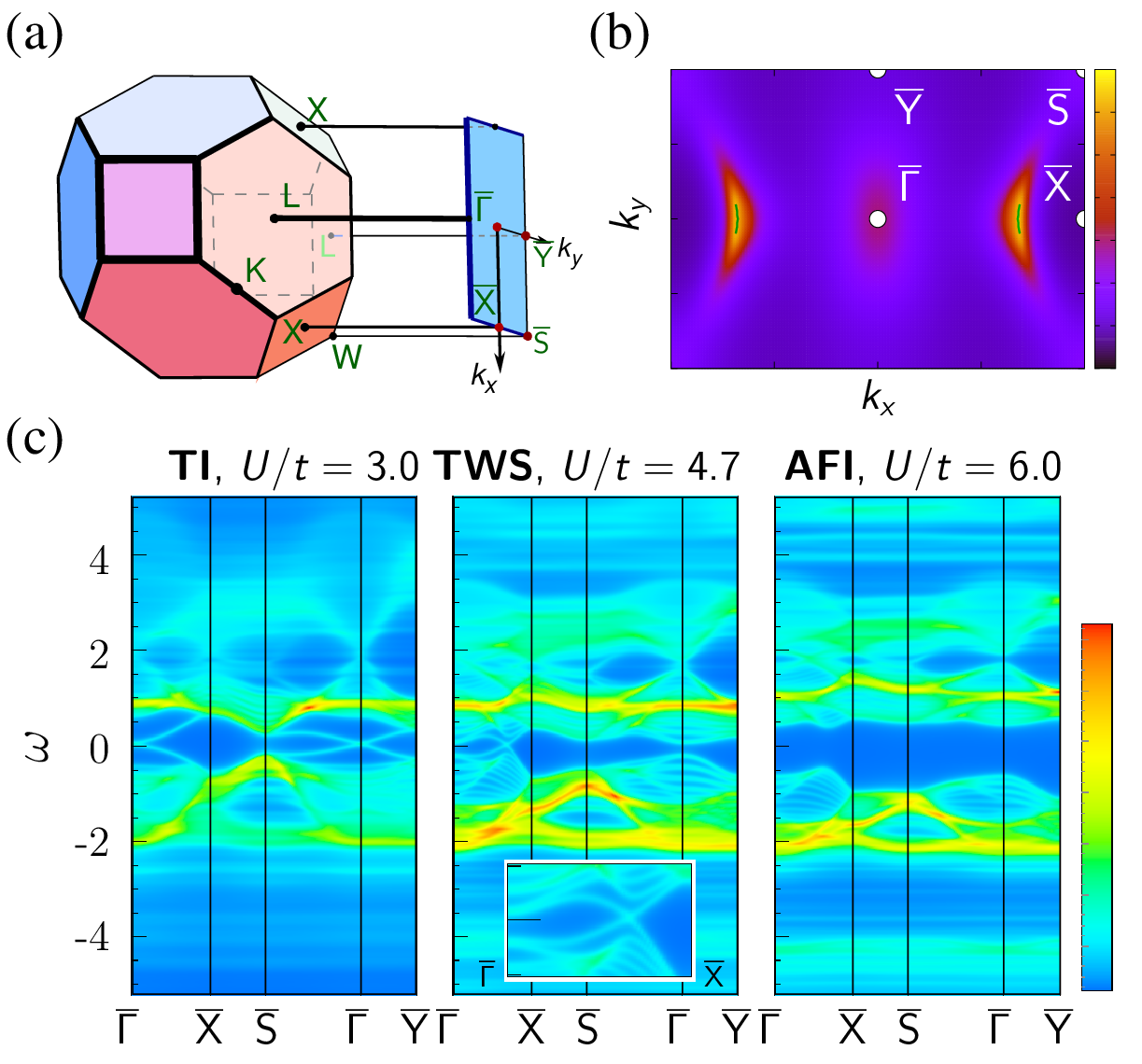}
	\caption{\label{fig.slab}%
	Surface states of a slab normal to (110).
	(a) First Brillouin zone (BZ) of a pyrochlore lattice
	and its projection onto the (110)-surface BZ.
	(b) 
        Spectral weight at the Fermi level, $\w=0$, for the TWS
	with $U/t=4.7$.
	Fermi arcs crossing $k_y=0$ clearly appear.
	The position of the arcs is roughly consistent
	with the HF result, denoted by green lines.
	(c) Density plot of the surface spectral functions
	along lines connecting high-symmetry points.
	From left to right, the panels represent a topological insulator (TI),
	a topological Weyl semimetal (TWS), and an antiferromagnetic insulator (AFI).
        The inset shows more clearly the surface states of the TWS near the Fermi level, $\w=0$.
	}
\end{figure}

\begin{figure}
	\includegraphics[width=1.01\linewidth]{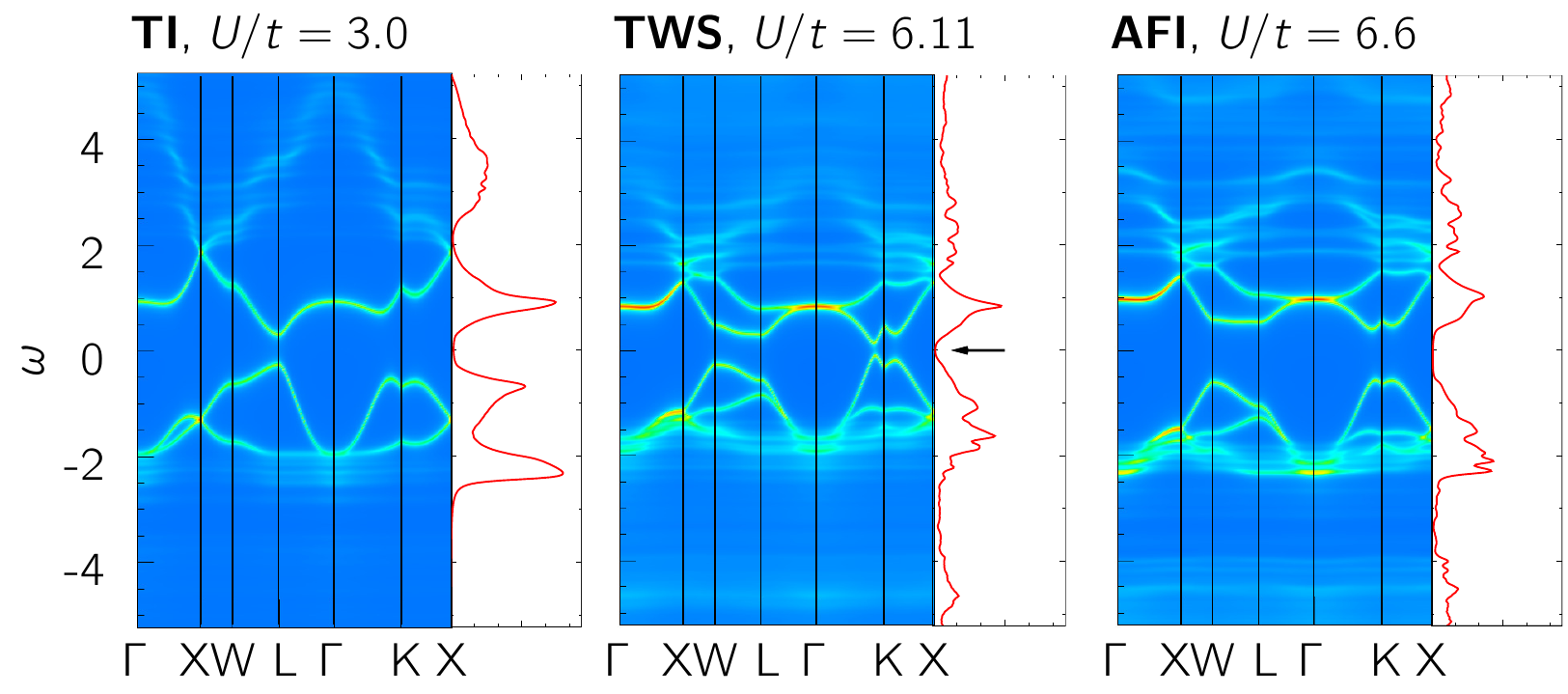}
	\caption{\label{fig.bulk_ts1}%
	Spectral weights along high symmetry lines
	and local density of states
	for different values of $U/t$.
	The panels represent a TI, a TWS, and an AF insulator.
	These last two phases break TRS. In the TWS, the Weyl
        points are not along high symmetry lines, but the density of states shows a quadratic
        scaling indicating their presence, as shown by the arrow.
	}
\end{figure}

The phase diagram together with the magnetization and topological index
are shown in Fig.~\ref{fig.phase}. After the magnetization jumps,
a topological Weyl semimetal emerges, as we establish from the spectral properties
of the surface (Fig.~\ref{fig.slab}) and bulk (Fig.~\ref{fig.bulk_ts1}) states.
The $\mathbb Z_2$ index, $\Delta$, 
 determines the presence of a quantized magneto-electric response. Specifically, 
$\D=1$ implies that an applied electric field $\mathbf E$ will induce a magnetization in a 
properly prepared system: $\b M=\al \b E$, where $\al=e^2/2h$ depends only on universal 
constants \cite{Hughes2008}. In the presence of TRS, this topological response can be used 
as a defining property of a correlated TI. The associated $\mathbb Z_2$ 
topological index can be computed from the full interacting Green's function by a 
Wess-Zumino-Witten like integral \cite{Wang2010}. It has been shown recently that 
in the special case where inversion symmetry is present,
as is the case in this work, one can use a simplified criterion~\cite{Wang2012}:
\begin{align}
  \label{eq:inv_short} 
  (-1)^\D = \prod_{\rm R-zero}\eta_{\al}^{1/2},
\end{align}
where $\eta_\al=\pm 1$ is a parity eigenvalue corresponding to a vector $\ket\al$,
an \emph{eigenstate of the interacting Green's function}
evaluated at one of eight special momenta, $\b\G_i$. 
These are the time reversal invariant momenta (TRIM) satisfying $-\b \G_i=\b \G_i$, up to a reciprocal lattice vector. 
\Req{inv_short} is in contrast with the analogous Fu-Kane formula which can only be used for non-interacting systems. 
More details about $\D$, such as the definition of ``R-zero'' (which reduces to that of an occupied
band in the non-interacting limit), 
can be found in the \methods~and in \cite{Wang2012}.

From Fig.~\ref{fig.phase}, we can see that the invariant indicates the presence of a topologically non-trivial phase
for a wide range of onsite repulsion until a trivial phase results in the magnetic antiferromagnet, found at large $U$. 
It can be noted that the topological index remains
invariant irrespective of the evolution of the Green's function due to interactions, which can be seen
from the broadening of the spectral function, for instance. 
Eventually, TRS is broken, and there is a quantum phase
transition out of the TI. From Fig.~\ref{fig.phase}, we note that there is a regime where 
the magnetization increases continuously 
from zero before jumping discontinuously at $U/t=6.11$. The latter jump,
where the order parameter has a sudden increase although TRS has already
been broken, signals a first order transition. We have verified that it is a robust property within
our framework. Figure~\ref{fig.phase} shows 
that the range where the magnetization increases continuoulsy from zero has $\D=1$. 
Because of the breaking of TRS, one cannot identify 
this as a TI in the above sense. 
Rather it is a closely-related phase: a \emph{correlated axion insulator}. It was introduced at the 
non-interacting level by Refs.~\cite{Turner2010,Wan2011},
where it was noted that even in the absence of TRS, by virtue of inversion symmetry and a
special structure of the parity eigenvalues, the topological magneto-electric effect 
discussed above could be
realized. Contrary to the TI, this phase does not have protected boundary states. 
As we argue in the \methods, the $\Delta$-invariant, \req{inv_short},
is a natural generalization of the one introduced in Refs. \cite{Turner2010,hughes2011} as it counts 
the total number of odd-parity eigenstates, not only one per Kramers pair. We add that
one expects such a phase to be present if the magnetization increases continuously from a
TI, because the parity structure is not expected to change dramatically.
Finally, as the continuous transition preceding the first order one
is a feature that is absent from the
HF mean-field theory, this axion phase is fundamentally correlation driven.

\emph{Surface states:} Another route to examining the non-trivial topology of the ground state 
is via the bulk-boundary correspondence
which, at the non-interacting level, guarantees the existence of protected surface states
on any boundary with a trivial insulator, such as the vacuum. We verify this correspondence at
the interacting level by performing a real space CDMFT calculation~\cite{Wu2011} on
a slab that is finite along one direction. 
We solve for
the layer-dependent Green's function self-consistently. The \methods~ contains 
details regarding the slab calculation.

The spectral function plotted in Fig.~\ref{fig.slab}(c) shows that the topological surface states persist 
as correlations are increased, the latter leading to spectral broadening and to the appearance of high energy states.
Eventually the slab system undergoes a first order transition to a TWS. At large $U$, we have an AFI
without any spectral weight in the gap coming from the boundaries.
Note that the
axion insulating phase presented in the above discussion for the bulk Hamiltonian is 
absent for the slab as there is no continuous rise of the magnetization. We attribute this to the finiteness of 
the system in one direction 
and expect the continuous transition to be recovered as one introduces more layers. 

\emph{Correlated topological Weyl semimetal:} 
After the magnetization jump in Fig.~\ref{fig.phase}, the spectral gap closes and one obtains a region of 
TWS before the AFI at large $U$.
The topological Weyl semimetal, as introduced at the non-interacting level \cite{Wan2011}, is a gapless
state with a Fermi surface consisting of (Weyl) points around which the dispersion is linear. 
These points are topologically robust as no local perturbation can gap them,
as long as two Weyl points of opposite chirality do not mix.
The protection of the Weyl node comes from the fact 
that only two bands meet at a point in three dimensions: all Pauli matrices have been used
in the Hamiltonian of the Weyl point
and an additional perturbation can only move the touching in the Brillouin zone (BZ). 
A fingerprint of the singular dispersion of the TWS
is that it harbors protected surface 
states which take the form of
Fermi arcs in the surface BZ \cite{Wan2011}.

At the level of the bulk calculation, we have determined that the spectral gap 
closes and the density of states shows a quadratic vanishing 
at the Fermi level ($\omega=0$), Fig.~\ref{fig.bulk_ts1}, as is expected from linearly dispersing fermions
in 3D. The eight Weyl points are not along high symmetry directions, hence Fig.~\ref{fig.bulk_ts1}
does not show the states of interest.
A decisive signature of the correlated Weyl phase comes from the surface state calculation.
We again consider a slab with surfaces normal to the $(110)$ direction. We find clear
Fermi arcs arising from states localized on the surfaces as we show in Fig.~\ref{fig.slab}(b). 
The arcs are broadened
compared to the sharp lines found at the non-interacting level. 
From the non-interacting 
theory we know that the arcs should join the projected bulk Weyl points on the surface BZ.
Moreover, if for a given surface two Weyl points of opposite chirality are projected onto each other, no
Fermi arc should arise from that point. In Fig.~\ref{fig.slab}(b), only two arcs can be seen for two reasons:
first, the arcs coming from the top and bottom surfaces overlap too much to be distinguished; second, two
pairs of opposite chirality are annihilated upon projection. We thus establish that the rule for the projection
holds in the correlated phase, i.e. the notion of chirality for the quasiparticles persists.

\emph{Discussion:}
We have so far mainly focused on theoretical studies of correlation effects on topological phases such as topological
insulators and Weyl semimetals.
Indeed, we have established the robustness of TIs from both sides of the
bulk-boundary duality. A bulk topological invariant defined in terms of interacting Green's functions was
explicitly evaluated. We determined its change at a correlation driven topological transition to a trivial
AF insulator. This invariant was used to predict the existence of a correlated axion phase at the onset
of a continuous magnetic transition. From the boundary perspective, our work has 
shown that the surface states of both TIs and 
topological Weyl semimetals remain robust to interactions. 

We now turn to the experimental considerations. The model we used is applicable to a large class
of complex oxides, the pyrochlore iridates. These show metal-insulator transitions 
as the rare earth is changed~\cite{Maeno2001} or pressure~\cite{Tafti2011} applied. 
There are indications that some members of the family
magnetically order at low temperatures~\cite{Zhao2011,Tomiyasu2011,Shapiro2012,Disseler2012,Qi2012}. 
However, it is still not clear what the nature of the ordering is, if any.
Diverse ground states can be realized as a result of the effects of chemical
and physical pressure on the electronic structure.
As correlations can play an important role in the determination of these ground states, 
it is important to understand their precise effect.
Our work goes beyond the non-interacting and mean-field studies done previously and establishes
not only the presence but also the stability of various topological phases and magnetic
orders with the inclusion of strong
correlations. Moreover, we predict that the axion insulator can in principle be realized due to the presence of a
correlation-driven second order transition preceding a first order one. In this phase, the surface states are gapped 
and the magneto-electric effect exists even though the bulk is magnetically ordered.
Generally, for both TI and axion phases, this suggests 
that a quantized magneto-electric response can be measured (by Kerr rotation
for example~\cite{Qi2011}) even if other probes, such as optical conductivity or photoemission,
point to the absence of sharp quasiparticles. It will be interesting to see if such indications
for correlated topological phases can be found in the iridates or other materials.
The methods used in our work, CDMFT (bulk and real space) and topological response
computed using Green's functions, can be used for a wide class of complex oxides, not only those
mentioned above. We suggest that these tools can be
applied to examine generic interacting states with symmetry-protected topological
order and combined with \emph{ab initio} tools in the quest for experimentally 
relevant candidate materials. 

We thank J.-M. Carter for his critical reading of the manuscript.
This work was supported by
the National Research Foundation of Korea (NRF) funded by the Korea government (MEST)
through the Quantum Metamaterials Research Center, No. 2011-0000982
and Basic Science Research, No. 2010-0010937 (AG, GSJ),
NSERC, the Canada Research Chair program, 
and the Canadian Institute for Advanced Research (WWK,YBK),
FQRNT and the Walter Sumner Foundation (WWK), NRF, No. 2008-0062238 (KP). 
The numerical computations were done in 
Seoul National University and SciNet at the University of Toronto.
\pagebreak
\onecolumngrid
\appendix
\section{Supplementary Material}\label{sec:methods}
\subsection{Cellular dynamical mean-field theory} 

\begin{figure}[h]
	\includegraphics[width=0.75\linewidth]{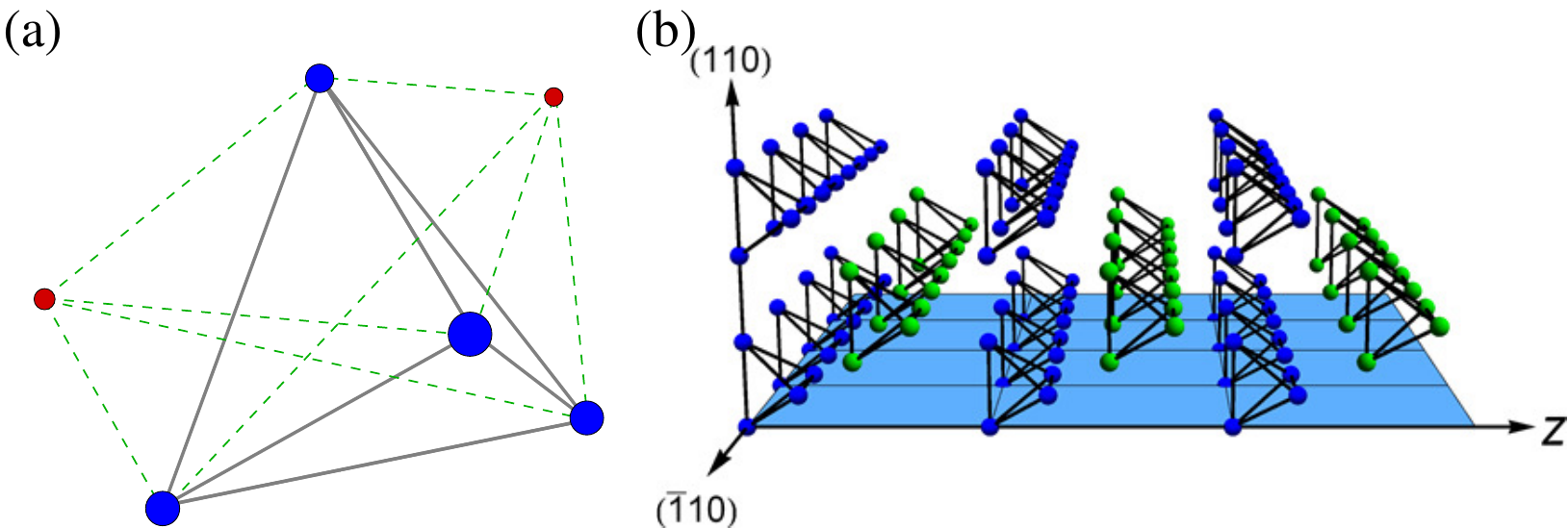}
	\caption{\label{fig.geo}%
	CDMFT cluster and slab geometry.
	(a) Unit tetrahedron (cluster) and bath sites. Only $N_b=2$ bath
        sites are shown	while $N_b=8~(4)$ were used for the bulk and slab calculations, respectively.
	The blue and red spheres denote the cluster and the bath sites, respectively.
	The dashed green lines indicate
	the effective hybridization between the cluster and the bath.
	(b) Three layers within a slab that is
          finite along the (110)-direction.
	The tetrahedra in each layer are colored alternately with blue or green.
	We use 16 layers with periodic boundary conditions for the other two directions
	which span the sky-blue plane.
	}
\end{figure} 

Cellular dynamical mean-field theory (CDMFT)~\cite{Kotliar2001,Maier2005} 
reduces infinite lattice to a cluster of size $N_c$
which hybridizes with the self-consistent electronic bath sites.
It extends single-site DMFT with the goal of capturing spatial correlations more adequately.
In this work, we use a tetrahedron cluster with 4 sites, which 
corresponds to a unit cell of the pyrochlore lattice.
Figure \ref{fig.geo} illustrates the cluster which is embedded in an effective, self-consistent bath.
In order to investigate the ground state properties,
we employ exact diagonalization to fully solve the quantum many-body
properties of the cluster.
The algorithm is iterative in nature:
we initially input an ansatz for the bath parameters and solve the hybridized Hamiltonian with the impurity solver.
From the cluster Hamiltonian we compute the cluster Green function $\hat{G}$,
the hat denoting an 8$\times$8 matrix structure,
as well as the cluster self-energy, $\hat{\Sigma}^c=\hat{\mathcal{G}}^{-1} - \hat{G}^{-1}$,
where $\hat{\mathcal{G}}$ is the Weiss field describing the noninteracting bath.
The new Weiss field is obtained by the self-consistent equation,
$\hat{\mathcal{G}}_\mathrm{new}^{-1} = \hat{G}^{-1}_{\mathrm{loc}}+ \hat{\Sigma}^c$,
where the local Green's function
\begin{align}
	\hat{G}_\mathrm{loc} (i\omega_n)
	= \sum_{ \tilde{\mathbf{k}} }\left[ (i\omega_n+\mu) \hat{1} - \hat{t}(\tilde{\mathbf{k}}) - \hat{\Sigma}^c(i\omega_n) \right]^{-1}
  \label{eq.Gk}
\end{align}
is calculated by integration over the momentum vector of the reduced Brillouin zone (BZ).
We then determine new bath parameters to best fit $\hat{\mathcal{G}}_\mathrm{new}$.
The steps are repeated until convergence is reached.

The calculation for a slab structure is similar to the bulk case,
but the Green's function is now an $8L \times 8L$ matrix.
The self-consistent equation for a slab with $L$ layers is given by
\begin{align}
	\bm{\mathcal{G}}^{-1}_{0,\mathrm{new}}(i\wn)
	= &\left[
	\sum_{k_x, k_y} \dfrac{1}{(i\wn + \mu) \mathbf{1} - \bm{t}(k_x,k_y) - \mathbf{\Sigma}(i\wn)}
	\right]^{-1} 
	+ \mathbf{\Sigma}(i\wn) \;,
\end{align}
where bold letters denote an $8L \times 8L$ matrix structure
and the summation runs over the surface Brillouin zone.
We assume that the self-energy is block-diagonal with the blocks of size eight.
The correlations beyond a unit cell are treated on the mean-field level.
In other words,
the sectors make an effect on each other via the hopping matrix $\bm{t}$,
although each sector of the self-energy is self-consistently determined by its own cluster Hamiltonian,
\begin{align}
  H^p_{\mathrm{c}} = & \nonumber
  \sum_{\mu\nu\sigma}E^p_{\mu\nu} c^{\dagger}_{p,\mu\sigma}c_{p,\nu\sigma}
  + U \sum_\mu n_{p,\mu\uparrow}n_{p,\mu\downarrow}
  \\ & 
  + \sum_{\mu l\sigma} ( V^{p}_{p,\mu l\sigma}  a^{\dagger}_{p,l\sigma}c^{}_{p,\mu\sigma} + {\rm h.c.} )
  + \sum_{l\sigma} \epsilon^{p}_{l\sigma}a^{\dagger}_{p,l\sigma} a^{}_{p,l\sigma},
\end{align}
where $\mu,\nu=1,2,\ldots,4$ are the site indices in a unit cell,
$l=1,2,\ldots,N_b$ label the bath sites,
and $p=1,\ldots,L$ is a layer index.
The hoppings within a cluster and chemical potential are introduced by $\hat{E}$
while the $L$-effective Weiss fields are described by the $V$'s and $\epsilon$'s.

\subsection{Topological invariant $\De$}

We provide information on the implementation of the topological invariant $\De$~\cite{Wang2012}
and its usage in the absence of TRS. For clarity, we rewrite it here,
\begin{align}
  \label{eq:inv_short2}
  (-1)^\D = \prod_{\textrm{R-zero}}\eta_{\al}^{1/2}\; .
\end{align}
As was noted in the main text, $\eta_\al=\pm 1$ is a parity eigenvalue corresponding 
to an eigenstate, $\ket\al$, of the (inverse) of the interacting Green's function, evaluated at one of eight TRIM, $\b\G_i$.
We now explain the notion of ``Right-zero'' (R-zero). The eigenvector satisfies:
\begin{align}
  \hat G\inv (i\w,\b \G_i) \ket{\al(\w,\b \G_i)} &= \mu_\al (\w,\b \G_i)\ket{\al(\w,\b \G_i)}\;, \\
  \hat P \ket{\al(\w,\b \G_i)} &= \eta_\al \ket{\al(\w,\b \G_i)}\;,
\end{align}
where $\hat G\inv(i\w,\b k)$ is the inverse Green's function, $\mu_\al(\w,\b k)$ is its generally
complex eigenvalue, and $\hat P$ the parity operator. For fixed $\b k$, as $\w$ is tuned 
from $-\infty$ to $\infty$, $\mu_\al(\w,\b k)$ sweeps a 
curve in the complex plane. 
At $\w=0$, $\hat G\inv(0,\b k)$ is Hermitian hence its eigenvalues are real. Thus, $\mu_\al(0,\b k)$
crosses the real axis in the complex plane and in this sense is called a ``zero''. It is a R-zero simply
if the crossing occurs to the right of the imaginary axis, i.e.
\begin{align}
  \textrm {R-zero} \Leftrightarrow \mu_\al(0,\b k) > 0\;.
\end{align}
Hence, if $\mu_\al(0,\b\G_i)$ is a R-zero, the associated parity eigenvalue $\eta_\al$ will contribute
to the product for $(-1)^\De$ above.

In the presence of TRS, each $\ket{\al(0,\b\G_i)}$ has a time-reversal partner with the same 
eigenvalue $\mu_\al$ \cite{Wang2012}. Each of these Kramers pairs of odd-parity eigenstates contributes $i^2=-1$,
and the index can only be zero or one. When one breaks TRS, the R-zeroes need not come in Kramers
pairs. However, we argue that even in that case, as long as inversion symmetry is preserved,
the topological index can still be used to test for the presence of a quantized magneto-electric response.
An insulator with TRS broken but with such a response is called an axion insulator \cite{Wan2011,Turner2010}.
The applicability of $\De$ for such inversion-symmetric topological insulators was briefly suggested 
in Ref.~\cite{Wang2012}. 

When TRS is broken, one needs to first ensure that $\De$ remains 0 or 1, avoiding imaginary values 
for instance. The equivalent statement for band insulators was established in 
references~\cite{Turner2010,hughes2011}, where it was shown that 
inversion symmetric band insulators always have an even number 
of odd-parity occupied states (including all TRIM). For correlated inversion-symmetric insulators, 
this would translate to the requirement to have an even number of odd-parity R-zeroes. This
property is expected for insulators that are adiabatically connected to a band-insulator. 
With this constraint, we can write an equivalent expression for $\De$:
\begin{align}
  \De = \frac{N_o}{2} \mod 2\;,
\end{align}
where $N_o$ is the total (even) number of odd-parity R-zeroes at the TRIM. This formula also applies to
the TRS case. In other words, for an insulator, a necessary condition to have
a topologically non-trivial magneto-electric response is that $N_o$ be twice an odd number. 
In the presence of TRS it is sufficient and we have the TI discussed above. Otherwise, one also needs 
to ascertain that the Hall conductivities vanish~\cite{Turner2010}. 

\begin{table}
\caption{%
Number of odd-parity {\it R-zeroes} per TRIM. 
The non-trivial Green's function
topology characterizes the topological insulator (TI) and axion insulator (AI)
phases. A topological transition occurs as the system enters the gapless TWS. The change
happens at the four $L$ points.
}\label{tbl:parities}
\begin{ruledtabular}
\begin{tabular}{ c c c c c c c }
	$U$ & Phase 
	&$\G$ & $\mathrm{X,Y,Z}$ & $\mathrm{L}^\prime$ & $\mathrm{L}(\times 3)$ & $N_o$ \\
\hline
0.00 & TI & 0 & 2 &   4 & 0  & 10\\
3.00 & TI & 0 & 2 &   4 & 0  & 10\\
6.00 & AI & 0 & 2 &   4 & 0  & 10\\
6.11 & TWS & 0 & 2 &   \bf 3 & \bf 1  & 12\\
8.00 & AFI & 0 & 2 &   \bf 3 & \bf 1  & 12
\end{tabular}
\end{ruledtabular}
\end{table}

We describe the evolution of the parities, and hence of $\De$, as we increase $U$ in the TI,
see Table~\ref{tbl:parities}. At $U=0$,
we have a topological band insulator and \req{inv_short2} reduces to the Fu-Kane index, $\nu_0$
\cite{Fu2007:2}.
Indeed, the condition of being a R-zero then implies an occupied Bloch state at the given TRIM. 
For $U>0$, we use the interacting Green's function to compute $\De$ and find that its parity
structure does not change as long as the state remains insulating.
This must be so as the gap does not close, hence none of the eigenvalues
of $\hat G\inv(0,\b\G_i)$ can vanish. Eventually, TRS is spontaneously broken but continuously.
For small values of the magnetization, not surprisingly, the parity structure is not affected and we
have an axion insulator. The magnetization jumps at the first order transition. 
The large magnetization after the transition alters  number of odd-parity
R-zeroes at the $L$ points such that $\De=0$ because $N_o=2\times 6$. 
Strictly speaking, one should not evaluate $\De$ near the transition as the system is gapless there, being 
in a TWS phase. Eventually, a gap opens leaving behind a trivial antiferromagnetic insulator (AFI). 

\bibliography{ref}{}

\begin{thebibliography}{30}%
\makeatletter
\providecommand \@ifxundefined [1]{%
 \@ifx{#1\undefined}
}%
\providecommand \@ifnum [1]{%
 \ifnum #1\expandafter \@firstoftwo
 \else \expandafter \@secondoftwo
 \fi
}%
\providecommand \@ifx [1]{%
 \ifx #1\expandafter \@firstoftwo
 \else \expandafter \@secondoftwo
 \fi
}%
\providecommand \natexlab [1]{#1}%
\providecommand \enquote  [1]{``#1''}%
\providecommand \bibnamefont  [1]{#1}%
\providecommand \bibfnamefont [1]{#1}%
\providecommand \citenamefont [1]{#1}%
\providecommand \href@noop [0]{\@secondoftwo}%
\providecommand \href [0]{\begingroup \@sanitize@url \@href}%
\providecommand \@href[1]{\@@startlink{#1}\@@href}%
\providecommand \@@href[1]{\endgroup#1\@@endlink}%
\providecommand \@sanitize@url [0]{\catcode `\\12\catcode `\$12\catcode
  `\&12\catcode `\#12\catcode `\^12\catcode `\_12\catcode `\%12\relax}%
\providecommand \@@startlink[1]{}%
\providecommand \@@endlink[0]{}%
\providecommand \url  [0]{\begingroup\@sanitize@url \@url }%
\providecommand \@url [1]{\endgroup\@href {#1}{\urlprefix }}%
\providecommand \urlprefix  [0]{URL }%
\providecommand \Eprint [0]{\href }%
\providecommand \doibase [0]{http://dx.doi.org/}%
\providecommand \selectlanguage [0]{\@gobble}%
\providecommand \bibinfo  [0]{\@secondoftwo}%
\providecommand \bibfield  [0]{\@secondoftwo}%
\providecommand \translation [1]{[#1]}%
\providecommand \BibitemOpen [0]{}%
\providecommand \bibitemStop [0]{}%
\providecommand \bibitemNoStop [0]{.\EOS\space}%
\providecommand \EOS [0]{\spacefactor3000\relax}%
\providecommand \BibitemShut  [1]{\csname bibitem#1\endcsname}%
\let\auto@bib@innerbib\@empty
\bibitem [{\citenamefont {Hasan}\ and\ \citenamefont {Kane}(2010)}]{Hasan2010}%
  \BibitemOpen
  \bibfield  {author} {\bibinfo {author} {\bibfnamefont {M.~Z.}\ \bibnamefont
  {Hasan}}\ and\ \bibinfo {author} {\bibfnamefont {C.~L.}\ \bibnamefont
  {Kane}},\ }\href {\doibase 10.1103/RevModPhys.82.3045} {\bibfield  {journal}
  {\bibinfo  {journal} {Rev. Mod. Phys.}\ }\textbf {\bibinfo {volume} {82}},\
  \bibinfo {pages} {3045} (\bibinfo {year} {2010})}\BibitemShut {NoStop}%
\bibitem [{\citenamefont {Qi}\ and\ \citenamefont {Zhang}(2011)}]{Qi2011}%
  \BibitemOpen
  \bibfield  {author} {\bibinfo {author} {\bibfnamefont {X.-L.}\ \bibnamefont
  {Qi}}\ and\ \bibinfo {author} {\bibfnamefont {S.-C.}\ \bibnamefont {Zhang}},\
  }\href {\doibase 10.1103/RevModPhys.83.1057} {\bibfield  {journal} {\bibinfo
  {journal} {Rev. Mod. Phys.}\ }\textbf {\bibinfo {volume} {83}},\ \bibinfo
  {pages} {1057} (\bibinfo {year} {2011})}\BibitemShut {NoStop}%
\bibitem [{\citenamefont {Turner}\ \emph
  {et~al.}(2010{\natexlab{a}})\citenamefont {Turner}, \citenamefont {Zhang},\
  and\ \citenamefont {Vishwanath}}]{Turner2010:2}%
  \BibitemOpen
  \bibfield  {author} {\bibinfo {author} {\bibfnamefont {A.~M.}\ \bibnamefont
  {Turner}}, \bibinfo {author} {\bibfnamefont {Y.}~\bibnamefont {Zhang}}, \
  and\ \bibinfo {author} {\bibfnamefont {A.}~\bibnamefont {Vishwanath}},\
  }\href {\doibase 10.1103/PhysRevB.82.241102} {\bibfield  {journal} {\bibinfo
  {journal} {Phys. Rev. B}\ }\textbf {\bibinfo {volume} {82}},\ \bibinfo
  {pages} {241102} (\bibinfo {year} {2010}{\natexlab{a}})}\BibitemShut
  {NoStop}%
\bibitem [{\citenamefont {Qi}\ \emph {et~al.}(2008)\citenamefont {Qi},
  \citenamefont {Hughes},\ and\ \citenamefont {Zhang}}]{Hughes2008}%
  \BibitemOpen
  \bibfield  {author} {\bibinfo {author} {\bibfnamefont {X.-L.}\ \bibnamefont
  {Qi}}, \bibinfo {author} {\bibfnamefont {T.~L.}\ \bibnamefont {Hughes}}, \
  and\ \bibinfo {author} {\bibfnamefont {S.-C.}\ \bibnamefont {Zhang}},\ }\href
  {\doibase 10.1103/PhysRevB.78.195424} {\bibfield  {journal} {\bibinfo
  {journal} {Phys. Rev. B}\ }\textbf {\bibinfo {volume} {78}},\ \bibinfo
  {pages} {195424} (\bibinfo {year} {2008})}\BibitemShut {NoStop}%
\bibitem [{\citenamefont {Wang}\ \emph {et~al.}(2010)\citenamefont {Wang},
  \citenamefont {Qi},\ and\ \citenamefont {Zhang}}]{Wang2010}%
  \BibitemOpen
  \bibfield  {author} {\bibinfo {author} {\bibfnamefont {Z.}~\bibnamefont
  {Wang}}, \bibinfo {author} {\bibfnamefont {X.-L.}\ \bibnamefont {Qi}}, \ and\
  \bibinfo {author} {\bibfnamefont {S.-C.}\ \bibnamefont {Zhang}},\ }\href
  {\doibase 10.1103/PhysRevLett.105.256803} {\bibfield  {journal} {\bibinfo
  {journal} {Phys. Rev. Lett.}\ }\textbf {\bibinfo {volume} {105}},\ \bibinfo
  {pages} {256803} (\bibinfo {year} {2010})}\BibitemShut {NoStop}%
\bibitem [{\citenamefont {Kotliar}\ \emph {et~al.}(2001)\citenamefont
  {Kotliar}, \citenamefont {Savrasov}, \citenamefont {P\'alsson},\ and\
  \citenamefont {Biroli}}]{Kotliar2001}%
  \BibitemOpen
  \bibfield  {author} {\bibinfo {author} {\bibfnamefont {G.}~\bibnamefont
  {Kotliar}}, \bibinfo {author} {\bibfnamefont {S.~Y.}\ \bibnamefont
  {Savrasov}}, \bibinfo {author} {\bibfnamefont {G.}~\bibnamefont {P\'alsson}},
  \ and\ \bibinfo {author} {\bibfnamefont {G.}~\bibnamefont {Biroli}},\
  }\href@noop {} {\bibfield  {journal} {\bibinfo  {journal} {Phys. Rev. Lett.}\
  }\textbf {\bibinfo {volume} {87}},\ \bibinfo {pages} {186401} (\bibinfo
  {year} {2001})}\BibitemShut {NoStop}%
\bibitem [{\citenamefont {Maier}\ \emph {et~al.}(2005)\citenamefont {Maier},
  \citenamefont {Jarrell}, \citenamefont {Pruschke},\ and\ \citenamefont
  {Hettler}}]{Maier2005}%
  \BibitemOpen
  \bibfield  {author} {\bibinfo {author} {\bibfnamefont {T.}~\bibnamefont
  {Maier}}, \bibinfo {author} {\bibfnamefont {M.}~\bibnamefont {Jarrell}},
  \bibinfo {author} {\bibfnamefont {T.}~\bibnamefont {Pruschke}}, \ and\
  \bibinfo {author} {\bibfnamefont {M.~H.}\ \bibnamefont {Hettler}},\
  }\href@noop {} {\bibfield  {journal} {\bibinfo  {journal} {Rev.\ Mod.\
  Phys.}\ }\textbf {\bibinfo {volume} {77}},\ \bibinfo {pages} {1027} (\bibinfo
  {year} {2005})}\BibitemShut {NoStop}%
\bibitem [{\citenamefont {Wan}\ \emph {et~al.}(2011)\citenamefont {Wan},
  \citenamefont {Turner}, \citenamefont {Vishwanath},\ and\ \citenamefont
  {Savrasov}}]{Wan2011}%
  \BibitemOpen
  \bibfield  {author} {\bibinfo {author} {\bibfnamefont {X.}~\bibnamefont
  {Wan}}, \bibinfo {author} {\bibfnamefont {A.~M.}\ \bibnamefont {Turner}},
  \bibinfo {author} {\bibfnamefont {A.}~\bibnamefont {Vishwanath}}, \ and\
  \bibinfo {author} {\bibfnamefont {S.~Y.}\ \bibnamefont {Savrasov}},\ }\href
  {\doibase 10.1103/PhysRevB.83.205101} {\bibfield  {journal} {\bibinfo
  {journal} {Phys. Rev. B}\ }\textbf {\bibinfo {volume} {83}},\ \bibinfo
  {pages} {205101} (\bibinfo {year} {2011})}\BibitemShut {NoStop}%
\bibitem [{\citenamefont {Balents}(2011)}]{Balents2011}%
  \BibitemOpen
  \bibfield  {author} {\bibinfo {author} {\bibfnamefont {L.}~\bibnamefont
  {Balents}},\ }\href {\doibase 10.1103/Physics.4.36} {\bibfield  {journal}
  {\bibinfo  {journal} {Physics}\ }\textbf {\bibinfo {volume} {4}},\ \bibinfo
  {pages} {36} (\bibinfo {year} {2011})}\BibitemShut {NoStop}%
\bibitem [{\citenamefont {Burkov}\ and\ \citenamefont
  {Balents}(2011)}]{Burkov2011}%
  \BibitemOpen
  \bibfield  {author} {\bibinfo {author} {\bibfnamefont {A.~A.}\ \bibnamefont
  {Burkov}}\ and\ \bibinfo {author} {\bibfnamefont {L.}~\bibnamefont
  {Balents}},\ }\href {\doibase 10.1103/PhysRevLett.107.127205} {\bibfield
  {journal} {\bibinfo  {journal} {Phys. Rev. Lett.}\ }\textbf {\bibinfo
  {volume} {107}},\ \bibinfo {pages} {127205} (\bibinfo {year}
  {2011})}\BibitemShut {NoStop}%
\bibitem [{\citenamefont {Yang}\ \emph {et~al.}(2011)\citenamefont {Yang},
  \citenamefont {Lu},\ and\ \citenamefont {Ran}}]{Yang2011}%
  \BibitemOpen
  \bibfield  {author} {\bibinfo {author} {\bibfnamefont {K.-Y.}\ \bibnamefont
  {Yang}}, \bibinfo {author} {\bibfnamefont {Y.-M.}\ \bibnamefont {Lu}}, \ and\
  \bibinfo {author} {\bibfnamefont {Y.}~\bibnamefont {Ran}},\ }\href {\doibase
  10.1103/PhysRevB.84.075129} {\bibfield  {journal} {\bibinfo  {journal} {Phys.
  Rev. B}\ }\textbf {\bibinfo {volume} {84}},\ \bibinfo {pages} {075129}
  (\bibinfo {year} {2011})}\BibitemShut {NoStop}%
\bibitem [{\citenamefont {Pesin}\ and\ \citenamefont
  {Balents}(2010)}]{Pesin2010}%
  \BibitemOpen
  \bibfield  {author} {\bibinfo {author} {\bibfnamefont {D.}~\bibnamefont
  {Pesin}}\ and\ \bibinfo {author} {\bibfnamefont {L.}~\bibnamefont
  {Balents}},\ }\href {\doibase 10.1038/nphys1606} {\bibfield  {journal}
  {\bibinfo  {journal} {Nature Phys.}\ }\textbf {\bibinfo {volume} {6}},\
  \bibinfo {pages} {376} (\bibinfo {year} {2010})}\BibitemShut {NoStop}%
\bibitem [{\citenamefont {Yang}\ and\ \citenamefont {Kim}(2010)}]{Yang2010}%
  \BibitemOpen
  \bibfield  {author} {\bibinfo {author} {\bibfnamefont {B.-J.}\ \bibnamefont
  {Yang}}\ and\ \bibinfo {author} {\bibfnamefont {Y.~B.}\ \bibnamefont {Kim}},\
  }\href {\doibase 10.1103/PhysRevB.82.085111} {\bibfield  {journal} {\bibinfo
  {journal} {Phys. Rev. B}\ }\textbf {\bibinfo {volume} {82}},\ \bibinfo
  {pages} {085111} (\bibinfo {year} {2010})}\BibitemShut {NoStop}%
\bibitem [{\citenamefont {Kargarian}\ \emph {et~al.}(2011)\citenamefont
  {Kargarian}, \citenamefont {Wen},\ and\ \citenamefont
  {Fiete}}]{Kargarian2011}%
  \BibitemOpen
  \bibfield  {author} {\bibinfo {author} {\bibfnamefont {M.}~\bibnamefont
  {Kargarian}}, \bibinfo {author} {\bibfnamefont {J.}~\bibnamefont {Wen}}, \
  and\ \bibinfo {author} {\bibfnamefont {G.~A.}\ \bibnamefont {Fiete}},\ }\href
  {\doibase 10.1103/PhysRevB.83.165112} {\bibfield  {journal} {\bibinfo
  {journal} {Phys. Rev. B}\ }\textbf {\bibinfo {volume} {83}},\ \bibinfo
  {pages} {165112} (\bibinfo {year} {2011})}\BibitemShut {NoStop}%
\bibitem [{\citenamefont {Witczak-Krempa}\ and\ \citenamefont
  {Kim}(2012)}]{Witczak-Krempa2011}%
  \BibitemOpen
  \bibfield  {author} {\bibinfo {author} {\bibfnamefont {W.}~\bibnamefont
  {Witczak-Krempa}}\ and\ \bibinfo {author} {\bibfnamefont {Y.~B.}\
  \bibnamefont {Kim}},\ }\href {\doibase 10.1103/PhysRevB.85.045124} {\bibfield
   {journal} {\bibinfo  {journal} {Phys. Rev. B}\ }\textbf {\bibinfo {volume}
  {85}},\ \bibinfo {pages} {045124} (\bibinfo {year} {2012})}\BibitemShut
  {NoStop}%
\bibitem [{\citenamefont {Witczak-Krempa}\ \emph {et~al.}(2010)\citenamefont
  {Witczak-Krempa}, \citenamefont {Choy},\ and\ \citenamefont {Kim}}]{tmi}%
  \BibitemOpen
  \bibfield  {author} {\bibinfo {author} {\bibfnamefont {W.}~\bibnamefont
  {Witczak-Krempa}}, \bibinfo {author} {\bibfnamefont {T.~P.}\ \bibnamefont
  {Choy}}, \ and\ \bibinfo {author} {\bibfnamefont {Y.~B.}\ \bibnamefont
  {Kim}},\ }\href {\doibase 10.1103/PhysRevB.82.165122} {\bibfield  {journal}
  {\bibinfo  {journal} {Phys. Rev. B}\ }\textbf {\bibinfo {volume} {82}},\
  \bibinfo {pages} {165122} (\bibinfo {year} {2010})}\BibitemShut {NoStop}%
\bibitem [{\citenamefont {Kotliar}\ \emph {et~al.}(2006)\citenamefont
  {Kotliar}, \citenamefont {Savrasov}, \citenamefont {Haule}, \citenamefont
  {Oudovenko}, \citenamefont {Parcollet},\ and\ \citenamefont
  {Marianetti}}]{Kotliar2006}%
  \BibitemOpen
  \bibfield  {author} {\bibinfo {author} {\bibfnamefont {G.}~\bibnamefont
  {Kotliar}}, \bibinfo {author} {\bibfnamefont {S.~Y.}\ \bibnamefont
  {Savrasov}}, \bibinfo {author} {\bibfnamefont {K.}~\bibnamefont {Haule}},
  \bibinfo {author} {\bibfnamefont {V.~S.}\ \bibnamefont {Oudovenko}}, \bibinfo
  {author} {\bibfnamefont {O.}~\bibnamefont {Parcollet}}, \ and\ \bibinfo
  {author} {\bibfnamefont {C.~A.}\ \bibnamefont {Marianetti}},\ }\href
  {\doibase 10.1103/RevModPhys.78.865} {\bibfield  {journal} {\bibinfo
  {journal} {Rev. Mod. Phys.}\ }\textbf {\bibinfo {volume} {78}},\ \bibinfo
  {pages} {865} (\bibinfo {year} {2006})}\BibitemShut {NoStop}%
\bibitem [{\citenamefont {Kim}\ \emph {et~al.}(2009)\citenamefont {Kim},
  \citenamefont {Ohsumi}, \citenamefont {Komesu}, \citenamefont {Sakai},
  \citenamefont {Morita}, \citenamefont {Takagi},\ and\ \citenamefont
  {Arima}}]{Kim2009}%
  \BibitemOpen
  \bibfield  {author} {\bibinfo {author} {\bibfnamefont {B.~J.}\ \bibnamefont
  {Kim}}, \bibinfo {author} {\bibfnamefont {H.}~\bibnamefont {Ohsumi}},
  \bibinfo {author} {\bibfnamefont {T.}~\bibnamefont {Komesu}}, \bibinfo
  {author} {\bibfnamefont {S.}~\bibnamefont {Sakai}}, \bibinfo {author}
  {\bibfnamefont {T.}~\bibnamefont {Morita}}, \bibinfo {author} {\bibfnamefont
  {H.}~\bibnamefont {Takagi}}, \ and\ \bibinfo {author} {\bibfnamefont
  {T.}~\bibnamefont {Arima}},\ }\href {\doibase 10.1126/science.1167106}
  {\bibfield  {journal} {\bibinfo  {journal} {Science}\ }\textbf {\bibinfo
  {volume} {323}},\ \bibinfo {pages} {1329} (\bibinfo {year}
  {2009})}\BibitemShut {NoStop}%
\bibitem [{\citenamefont {Wu}\ \emph {et~al.}(2011)\citenamefont {Wu},
  \citenamefont {Rachel}, \citenamefont {Liu},\ and\ \citenamefont {{Le
  H}ur}}]{Wu2011}%
  \BibitemOpen
  \bibfield  {author} {\bibinfo {author} {\bibfnamefont {W.}~\bibnamefont
  {Wu}}, \bibinfo {author} {\bibfnamefont {S.}~\bibnamefont {Rachel}}, \bibinfo
  {author} {\bibfnamefont {W.-M.}\ \bibnamefont {Liu}}, \ and\ \bibinfo
  {author} {\bibfnamefont {K.}~\bibnamefont {{Le H}ur}},\ }\href@noop {} {\
  (\bibinfo {year} {2011})},\ \Eprint {http://arxiv.org/abs/1106.0943}
  {arXiv:1106.0943} \BibitemShut {NoStop}%
\bibitem [{\citenamefont {Wang}\ \emph {et~al.}(2012)\citenamefont {Wang},
  \citenamefont {Qi},\ and\ \citenamefont {Zhang}}]{Wang2012}%
  \BibitemOpen
  \bibfield  {author} {\bibinfo {author} {\bibfnamefont {Z.}~\bibnamefont
  {Wang}}, \bibinfo {author} {\bibfnamefont {X.-L.}\ \bibnamefont {Qi}}, \ and\
  \bibinfo {author} {\bibfnamefont {S.-C.}\ \bibnamefont {Zhang}},\ }\href@noop
  {} {\  (\bibinfo {year} {2012})},\ \Eprint {http://arxiv.org/abs/1201.6431v2}
  {arXiv:1201.6431v2} \BibitemShut {NoStop}%
\bibitem [{\citenamefont {Turner}\ \emph
  {et~al.}(2010{\natexlab{b}})\citenamefont {Turner}, \citenamefont {Yi},
  \citenamefont {Mong},\ and\ \citenamefont {Vishwanath}}]{Turner2010}%
  \BibitemOpen
  \bibfield  {author} {\bibinfo {author} {\bibfnamefont {A.~M.}\ \bibnamefont
  {Turner}}, \bibinfo {author} {\bibfnamefont {Z.}~\bibnamefont {Yi}}, \bibinfo
  {author} {\bibfnamefont {R.~S.~K.}\ \bibnamefont {Mong}}, \ and\ \bibinfo
  {author} {\bibfnamefont {A.}~\bibnamefont {Vishwanath}},\ }\href@noop {} {\
  (\bibinfo {year} {2010}{\natexlab{b}})},\ \Eprint
  {http://arxiv.org/abs/1010.4335v2} {arXiv:1010.4335v2} \BibitemShut {NoStop}%
\bibitem [{\citenamefont {Hughes}\ \emph {et~al.}(2011)\citenamefont {Hughes},
  \citenamefont {Prodan},\ and\ \citenamefont {Bernevig}}]{hughes2011}%
  \BibitemOpen
  \bibfield  {author} {\bibinfo {author} {\bibfnamefont {T.~L.}\ \bibnamefont
  {Hughes}}, \bibinfo {author} {\bibfnamefont {E.}~\bibnamefont {Prodan}}, \
  and\ \bibinfo {author} {\bibfnamefont {B.~A.}\ \bibnamefont {Bernevig}},\
  }\href {\doibase 10.1103/PhysRevB.83.245132} {\bibfield  {journal} {\bibinfo
  {journal} {Phys. Rev. B}\ }\textbf {\bibinfo {volume} {83}},\ \bibinfo
  {pages} {245132} (\bibinfo {year} {2011})}\BibitemShut {NoStop}%
\bibitem [{\citenamefont {Yanagishima}\ and\ \citenamefont
  {Maeno}(2001)}]{Maeno2001}%
  \BibitemOpen
  \bibfield  {author} {\bibinfo {author} {\bibfnamefont {D.}~\bibnamefont
  {Yanagishima}}\ and\ \bibinfo {author} {\bibfnamefont {Y.}~\bibnamefont
  {Maeno}},\ }\href {http://jpsj.ipap.jp/link?JPSJ/70/2880/} {\bibfield
  {journal} {\bibinfo  {journal} {J. Phys. Soc. Jpn.}\ }\textbf {\bibinfo
  {volume} {70}},\ \bibinfo {pages} {2880} (\bibinfo {year}
  {2001})}\BibitemShut {NoStop}%
\bibitem [{\citenamefont {{Tafti}}\ \emph {et~al.}(2011)\citenamefont
  {{Tafti}}, \citenamefont {{Ishikawa}}, \citenamefont {{McCollam}},
  \citenamefont {{Nakatsuji}},\ and\ \citenamefont {{Julian}}}]{Tafti2011}%
  \BibitemOpen
  \bibfield  {author} {\bibinfo {author} {\bibfnamefont {F.~F.}\ \bibnamefont
  {{Tafti}}}, \bibinfo {author} {\bibfnamefont {J.~J.}\ \bibnamefont
  {{Ishikawa}}}, \bibinfo {author} {\bibfnamefont {A.}~\bibnamefont
  {{McCollam}}}, \bibinfo {author} {\bibfnamefont {S.}~\bibnamefont
  {{Nakatsuji}}}, \ and\ \bibinfo {author} {\bibfnamefont {S.~R.}\ \bibnamefont
  {{Julian}}},\ }\href@noop {} {\  (\bibinfo {year} {2011})},\ \Eprint
  {http://arxiv.org/abs/1107.2544} {arXiv:1107.2544} \BibitemShut {NoStop}%
\bibitem [{\citenamefont {Zhao}\ \emph {et~al.}(2011)\citenamefont {Zhao},
  \citenamefont {Mackie}, \citenamefont {MacLaughlin}, \citenamefont {Bernal},
  \citenamefont {Ishikawa}, \citenamefont {Ohta},\ and\ \citenamefont
  {Nakatsuji}}]{Zhao2011}%
  \BibitemOpen
  \bibfield  {author} {\bibinfo {author} {\bibfnamefont {S.}~\bibnamefont
  {Zhao}}, \bibinfo {author} {\bibfnamefont {J.~M.}\ \bibnamefont {Mackie}},
  \bibinfo {author} {\bibfnamefont {D.~E.}\ \bibnamefont {MacLaughlin}},
  \bibinfo {author} {\bibfnamefont {O.~O.}\ \bibnamefont {Bernal}}, \bibinfo
  {author} {\bibfnamefont {J.~J.}\ \bibnamefont {Ishikawa}}, \bibinfo {author}
  {\bibfnamefont {Y.}~\bibnamefont {Ohta}}, \ and\ \bibinfo {author}
  {\bibfnamefont {S.}~\bibnamefont {Nakatsuji}},\ }\href {\doibase
  10.1103/PhysRevB.83.180402} {\bibfield  {journal} {\bibinfo  {journal} {Phys.
  Rev. B}\ }\textbf {\bibinfo {volume} {83}},\ \bibinfo {pages} {180402}
  (\bibinfo {year} {2011})}\BibitemShut {NoStop}%
\bibitem [{\citenamefont {Tomiyasu}\ \emph {et~al.}(2011)\citenamefont
  {Tomiyasu}, \citenamefont {Matsuhira}, \citenamefont {Iwasa}, \citenamefont
  {Watahiki}, \citenamefont {Takagi}, \citenamefont {Wakeshima}, \citenamefont
  {Hinatsu}, \citenamefont {Yokoyama}, \citenamefont {Ohoyama},\ and\
  \citenamefont {Yamada}}]{Tomiyasu2011}%
  \BibitemOpen
  \bibfield  {author} {\bibinfo {author} {\bibfnamefont {K.}~\bibnamefont
  {Tomiyasu}}, \bibinfo {author} {\bibfnamefont {K.}~\bibnamefont {Matsuhira}},
  \bibinfo {author} {\bibfnamefont {K.}~\bibnamefont {Iwasa}}, \bibinfo
  {author} {\bibfnamefont {M.}~\bibnamefont {Watahiki}}, \bibinfo {author}
  {\bibfnamefont {S.}~\bibnamefont {Takagi}}, \bibinfo {author} {\bibfnamefont
  {M.}~\bibnamefont {Wakeshima}}, \bibinfo {author} {\bibfnamefont
  {Y.}~\bibnamefont {Hinatsu}}, \bibinfo {author} {\bibfnamefont
  {M.}~\bibnamefont {Yokoyama}}, \bibinfo {author} {\bibfnamefont
  {K.}~\bibnamefont {Ohoyama}}, \ and\ \bibinfo {author} {\bibfnamefont
  {K.}~\bibnamefont {Yamada}},\ }\href@noop {} {\  (\bibinfo {year} {2011})},\
  \Eprint {http://arxiv.org/abs/1110.6605} {arXiv:1110.6605} \BibitemShut
  {NoStop}%
\bibitem [{\citenamefont {Shapiro}\ \emph {et~al.}(2012)\citenamefont
  {Shapiro}, \citenamefont {Riggs}, \citenamefont {Stone}, \citenamefont
  {Cruz}, \citenamefont {Chi}, \citenamefont {Podlesnyak},\ and\ \citenamefont
  {Fisher}}]{Shapiro2012}%
  \BibitemOpen
  \bibfield  {author} {\bibinfo {author} {\bibfnamefont {M.~C.}\ \bibnamefont
  {Shapiro}}, \bibinfo {author} {\bibfnamefont {S.~C.}\ \bibnamefont {Riggs}},
  \bibinfo {author} {\bibfnamefont {M.~B.}\ \bibnamefont {Stone}}, \bibinfo
  {author} {\bibfnamefont {C.~R. d.~l.}\ \bibnamefont {Cruz}}, \bibinfo
  {author} {\bibfnamefont {S.}~\bibnamefont {Chi}}, \bibinfo {author}
  {\bibfnamefont {A.~A.}\ \bibnamefont {Podlesnyak}}, \ and\ \bibinfo {author}
  {\bibfnamefont {I.~R.}\ \bibnamefont {Fisher}},\ }\href@noop {} {\  (\bibinfo
  {year} {2012})},\ \Eprint {http://arxiv.org/abs/1201.5419} {arXiv:1201.5419}
  \BibitemShut {NoStop}%
\bibitem [{\citenamefont {Disseler}\ \emph {et~al.}(2012)\citenamefont
  {Disseler}, \citenamefont {Dhital}, \citenamefont {Hogan}, \citenamefont
  {Amato}, \citenamefont {Giblin}, \citenamefont {Cruz}, \citenamefont
  {{Daoud-Aladine}}, \citenamefont {Wilson},\ and\ \citenamefont
  {Graf}}]{Disseler2012}%
  \BibitemOpen
  \bibfield  {author} {\bibinfo {author} {\bibfnamefont {S.~M.}\ \bibnamefont
  {Disseler}}, \bibinfo {author} {\bibfnamefont {C.}~\bibnamefont {Dhital}},
  \bibinfo {author} {\bibfnamefont {T.~C.}\ \bibnamefont {Hogan}}, \bibinfo
  {author} {\bibfnamefont {A.}~\bibnamefont {Amato}}, \bibinfo {author}
  {\bibfnamefont {S.~R.}\ \bibnamefont {Giblin}}, \bibinfo {author}
  {\bibfnamefont {C.~d.}\ \bibnamefont {Cruz}}, \bibinfo {author}
  {\bibfnamefont {A.}~\bibnamefont {{Daoud-Aladine}}}, \bibinfo {author}
  {\bibfnamefont {S.~D.}\ \bibnamefont {Wilson}}, \ and\ \bibinfo {author}
  {\bibfnamefont {M.~J.}\ \bibnamefont {Graf}},\ }\href@noop {} {\  (\bibinfo
  {year} {2012})},\ \Eprint {http://arxiv.org/abs/1201.4606} {arXiv:1201.4606}
  \BibitemShut {NoStop}%
\bibitem [{\citenamefont {Qi}\ \emph {et~al.}(2012)\citenamefont {Qi},
  \citenamefont {Korneta}, \citenamefont {Wan},\ and\ \citenamefont
  {Cao}}]{Qi2012}%
  \BibitemOpen
  \bibfield  {author} {\bibinfo {author} {\bibfnamefont {T.~F.}\ \bibnamefont
  {Qi}}, \bibinfo {author} {\bibfnamefont {O.~B.}\ \bibnamefont {Korneta}},
  \bibinfo {author} {\bibfnamefont {X.}~\bibnamefont {Wan}}, \ and\ \bibinfo
  {author} {\bibfnamefont {G.}~\bibnamefont {Cao}},\ }\href@noop {} {\
  (\bibinfo {year} {2012})},\ \Eprint {http://arxiv.org/abs/1201.0538}
  {arXiv:1201.0538} \BibitemShut {NoStop}%
\bibitem [{\citenamefont {Fu}\ and\ \citenamefont {Kane}(2007)}]{Fu2007:2}%
  \BibitemOpen
  \bibfield  {author} {\bibinfo {author} {\bibfnamefont {L.}~\bibnamefont
  {Fu}}\ and\ \bibinfo {author} {\bibfnamefont {C.~L.}\ \bibnamefont {Kane}},\
  }\href {\doibase 10.1103/PhysRevB.76.045302} {\bibfield  {journal} {\bibinfo
  {journal} {Phys. Rev. B}\ }\textbf {\bibinfo {volume} {76}},\ \bibinfo
  {pages} {045302} (\bibinfo {year} {2007})}\BibitemShut {NoStop}%
\end{thebibliography}%
\end{document}